\RequirePackage{amsthm}
\documentclass[sn-mathphys,Numbered]{sn-jnl}

\usepackage{graphicx}%
\usepackage{multirow}%
\usepackage{amsmath,amssymb,amsfonts}%
\usepackage{natbib}

\usepackage{amsthm}%
\usepackage{mathrsfs}%
\usepackage[title]{appendix}%
\usepackage{xcolor}%
\usepackage{textcomp}%
\usepackage{manyfoot}%
\usepackage{booktabs}%
\usepackage{algorithm}%
\usepackage{algorithmicx}%
\usepackage{algpseudocode}%
\usepackage{listings}%
\usepackage{float}
\usepackage{subfigure}

\theoremstyle{thmstyleone}%

\theoremstyle{thmstyletwo}%

\theoremstyle{thmstylethree}%

\begin{document}

\title[Article Title]{The Emergence of Cooperation in the well-mixed Prisoner's Dilemma: Memory Couples Individual and Group Strategies}


\author[1]{\fnm{Changyan} \sur{Di}}\email{dizhy@lzu.edu.cn}

\author[2]{\fnm{Jianyue} \sur{Guan}}\email{guanjy@lzu.edu.cn}

\author*[1]{\fnm{Qingguo} \sur{Zhou}}\email{zhouqg@lzu.edu.cn}
\author[1]{\fnm{Jingqiang} \sur{Wang}}\email{jqwang16@lzu.edu.cn}
\author[1]{\fnm{Xiangyang} \sur{Li}}\email{lixiangyang20@lzu.edu.cn}

\affil[1]{\orgdiv{Lanzhou University}, \orgname{School of information science and technology}, \orgaddress{\street{Tianshui Road}, \city{lanzhou}, \postcode{730000}, \state{Gansu Province}, \country{China}}}

\affil[2]{\orgdiv{Department}, \orgname{Organization}, \orgaddress{\street{Street}, \city{City}, \postcode{10587}, \state{State}, \country{Country}}}


\abstract{
Exploration of mechanisms underlying the emergence of collective cooperation remains a focal point in field of evolution of cooperation. Prevailing studies often neglect historical information, relying on the latest rewards as the primary criterion for individual decision-making—a method incongruent with human cognition and decision-making modes. This limitation impedes a comprehensive understanding of the spontaneous emergence of cooperation. Integrating memory factors into evolutionary game models to formulate decision criteria with delayed effects has shown potential in unraveling cooperation mechanisms. However, this comes at the significant cost of heightened computational complexity. In this paper, we propose an experiential decision-making method based on reinforcement learning. Utilizing this method, we construct a multi-agent system to engage in the evolutionary Prisoner's Dilemma game. Simulation results indicate that memory establishes a coupling relationship between individual and group strategies, fostering periodic oscillation between cooperation and defection in a well-mixed group. Specifically, defection loses its payoff advantage over cooperation as the group cooperation rate decreases. Conversely, the cooperative behavior gains reinforcement with an increase in the group cooperation rate, overcoming defection as the dominant strategy for individuals. This coupling between individual and group strategies fundamentally bridges the gap between individual and group interests, integrating a multitude of known factors and elucidating the fundamental mechanism of cooperation emergence in the face of social dilemmas.}

\keywords{Prisoner's Dilemma game, reinforcement learning, emergence of collective cooperation, memory}



\maketitle
\section{Introduction}\label{sec1}

Conflicts between individual interests and collective interests are pervasive, as exemplified by the recent closure of COP28, aimed at fostering global cooperation to address climate change\cite{WEF_COP28}. The challenge lies in orchestrating global collaboration among countries or entities facing the dilemma of developing their own economies versus prioritizing the overall human interest in the absence of binding constraints. Game theory serves as a potent tool for comprehending conflicts and cooperation among individuals\cite{smith1973logic}. Three prototypical models have been proposed to describe distinct cooperation dilemmas encountered in human society: the Prisoner's Dilemma game (PDG)\cite{rapoport1965prisoner}, the Snowdrift game\cite{smith1973logic,sugden2004economics}, and the Stag Hunt game\cite{skyrms2004stag}. 
Evolutionary game theory extends game-theoretic principles to model dynamic changes in the frequency of strategists\cite{smith1982evolution,nowak2004emergence}, providing a unified framework for studying the three aforementioned models. Unlike the latter two models that allow cooperation and defection to coexist in well-mixed groups, within the framework of evolutionary game theory, defection stands as a strict Nash equilibrium point in the PDG. Consequently, the evolution of well-mixed groups inevitably leads to a situation where cooperators are overwhelmingly ousted\cite{doebeli2005models,axelrod1981evolution}.

In order to explicate the ubiquitous phenomenon of cooperation in reality, a mechanism for the evolution of cooperation is often necessary\cite{nowak2006five}. Thousands of studies have provided valuable insights into the motivations (e.g., self-interest, reciprocity, fairness)\cite{nowak2005evolution,bicchieri2010behaving,henrich2010markets,hughes2018inequity}, institutions (sanctioning and reputation systems, social norms), structure\cite{szolnoki2009topology,nowak1992evolutionary}, and dynamics (e.g., conditional cooperation) associated with human cooperation\cite{otten2022human}. Recent researches has spotlighted evolutionary models with environmental feedback\cite{weitz2016oscillating,tilman2020evolutionary,szolnoki2018environmental} and multi-modal games\cite{hilbe2018evolution,wang2014different,szolnoki2014coevolutionary}, emphasizing the variability of the payoff matrix across different situations. Despite the plethora of proposed mechanisms aimed at facilitating cooperative evolution, their applicability varies across different scenarios and entities. For instance, cooperation observed among animals or microorganisms eludes a sole explanation through human-specific emotional factors\cite{wenseleers2004tragedy,falster2003plant,foster2004diminishing}. Kin selection can only account for groups with genetic relatedness, and the gain advantage of cooperators in environmental feedback is artificially set under specific environmental conditions\cite{weitz2016oscillating}. Therefore, for the explanation of the emergence and evolution of group cooperation, is there a unified mechanism? What is this fundamental one? These inquiries underscore the imperative for further in-depth research in the field of the emergence mechanism of group cooperation.

In the investigation of cooperative evolution mechanisms, the updating methods of individual strategies play a crucial role. The most commonly adopted methods in existing literature are imitator rules and replicator equations\cite{taylor1978evolutionary}. The former is based on Monte Carlo simulation methods, while the latter is an analytical approach by mathematical theory. However, the fundamental assumption of both methods is that individuals are memoryless and myopic\cite{hilbe2017memory,qin2008effect}, that they make decisions based on the recent round's rewards\cite{wang2016cooperation}. This premise contradicts findings in related fields. For example, researches in cognitive and behavioral neuroscience have shown that integrated memories are invoked during novel situations to facilitate various behaviors, from spatial navigation to imagination\cite{schlichting2015memory}. And memory plays an essential role in a wide range of value-based decisions\cite{biderman2020memories}. Moreover, in the field of behavioral experiments, Burton-Chellew et al. demonstrates that individuals largely act as self-interested ``confused learners'' in public goods games, starting with some imperfect idea of maximizing their own payoff and learning from experience how their contribution affects their payoff\cite{burton2021payoff}. McAuliffe et al. argue that individuals accumulate experience and adjust cooperative behavior based on payoff situations when faced with unfamiliar situations\cite{mcauliffe2019cooperation,burton2016conditional}. 

Recently, there have been numerous studies that introduce memory into evolutionary game models. For example, C. Hilbe employed an axiomatic approach to analyze the characteristics of robust cooperative strategies, highlighting the value of memory-n strategies when the benefit of cooperation is small or intermediate\cite{hilbe2017memory}. Liu et al., based on network structure, investigated the effects of memory proportion, memory factor, memory length, and adaptive sensitivity index on group cooperation\cite{liu2010memory,wang2006memory,shu2019memory,deng2021memory,ye2017memory,stewart2016small}. These studies, from various perspectives, analyzed the role and effects of memory mechanisms in evolutionary games, reaching a general conclusion that the introduction of memory can effectively enhance cooperation in various scenarios. This demonstrates the potential of memory to deepen the understanding of potential cooperation mechanisms. However, it's worth noting that most of the aforementioned studies are built on spatial structures and have not been conducted in well-mixed groups. Furthermore, introducing memory into the well-mixed group poses a challenge. Memory involves storing information about one's own strategies and the strategies and payoffs of opponents. This significantly increases the complexity of computation to the extent that it becomes impractical, especially in cases of large population size\cite{hilbe2017memory}. Some literature has explored mixed homogeneous groups adopting the Q-learning method. For instance, Wang et al. studied the prisoner's dilemma game with reinforcement learning, revealing a particular positive role of Levy noise in the evolutionary dynamics of social dilemmas\cite{wang2022levy}. Zhang et al. applied the Q-learning algorithm to agents seeking optimal actions based on introspection principles in a well-mixed system, providing inspiration for the study of cooperative phenomena and oscillatory behaviors\cite{zhang2020oscillatory}. However, Q-learning in its original form, is better suited for exploring sequential decision problems within discrete state spaces in static environments. It is not well-suited for group games where the environment is dynamic, shaped by the collaborative actions of a group that are challenging to define, and may lead to potential state explosions\cite{sandholm1996multiagent,kies2020finding}.

Therefore, this paper has modified the classical Q-learning method and devised an Individual Experiential Decision-Making (EDM) approach that incorporates memory to formulate a decision criterion with delayed effects. Utilizing this method, a multi-agent system (MAS) model was constructed to delve into the fundamental mechanisms underlying the emergence of cooperation in an evolutionary PDG with well-mixed population from a bottom-up perspective.

Simulation results demonstrate that the introduction of memory establishes a coupling relationship between individual and group strategies, leading to periodic oscillations between cooperation and defection. Specifically, when individuals pursue higher payoffs by adopting a defection strategy, this coupling relationship gradually increases the number of defectors in the group, causing a decline in their experiential expected payoffs and a loss of relative advantage over cooperation strategies. Greedy individuals are compelled to alternate between different strategies in search of higher payoffs. During this process, when the majority of individuals in the group synchronously flip from defectors to cooperators, cooperation strategies stand out due to the reinforcement of immediate payoffs. Group cooperation emerges from the disorder state. However, the inherent greed of individuals leads to the collapse of group cooperation over time, resulting in repetitive cyclical oscillations.

The Experiential Decision-Making (EDM) method proposed in this study eliminates the necessity for individual memory regarding historical payoffs and strategies. Instead, it concentrates on learning the associative relationships between them, leading to a substantial reduction in the algorithm's complexity. The uncovered coupling relationship between individual and group strategies, under this approach, fundamentally closes the divide between individual and group interests. This revelation elucidates the fundamental mechanism behind the emergence of cooperation in the context of social dilemmas.


\section{Model and Methods}\label{sec2}

We have developed a Multi-Agent System (MAS) for the evolutionary PDG involving a population of $N$ individuals. In this model, each agent follows an identical strategy set $\mathcal{S}$, learning methodology, and decision-making algorithm. Considering a symmetric two-player game, where players can choose between cooperation(denoted as $C$) and defection(denoted as $D$). In this game, mutual cooperation results in a reward payoff of $R$, mutual defection leads to a punishment payoff of $P$, while mixed choices give rise to the sucker's payoff $S$ for the cooperator and the tempting payoff $T$ for the defector. For simplicity, we set $R = 1$ and $P = 0$, and specify the other payoffs as $T=b$ and $S=-c$ (where $c \geq 0$). This configuration is visually depicted in Fig.\ref{fig:model}.

In evolutionary games, Q-learning stands out as the prevailing reinforcement learning method. However, classical Q-learning is particularly adept at addressing sequential decision problems, necessitating the modeling of the state space. Consequently, it finds common application in tasks characterized by static environments featuring finite and well-defined state spaces, such as maze-solving or two-player games. In such scenarios, the Q-table tends to converge through iterative updates. In contrast, in group games, the environment undergoes dynamic changes with individual actions. When the number of agents is excessively high, this dynamism may result in a state space explosion \cite{sandholm1996multiagent, kies2020finding}, significantly diminishing the applicability of Q-learning. Moreover, it is crucial to emphasize that, in Q-learning, the term "state" pertains to the environmental state, not the state of individual agents. Some studies consider the actions of agents as their states, which is clearly at odds with the intended purpose of the model. In fact, in evolutionary games, individuals focus on strategies with long-term stable payoffs, making the multi-armed bandit model more suitable than Q-learning in such scenarios. This model disregards states and concentrates on balancing actions to maximize the expected value of cumulative rewards.
Therefore, this paper introduces a novel approach by integrating the Q-learning algorithm and the multi-armed bandit algorithm, devising an experiential decision-making method that incorporates memory to form criteria with delayed effects. The learning method is as follows:

\begin{figure*}[tp]
    \centering{\includegraphics[width=0.7\textwidth]{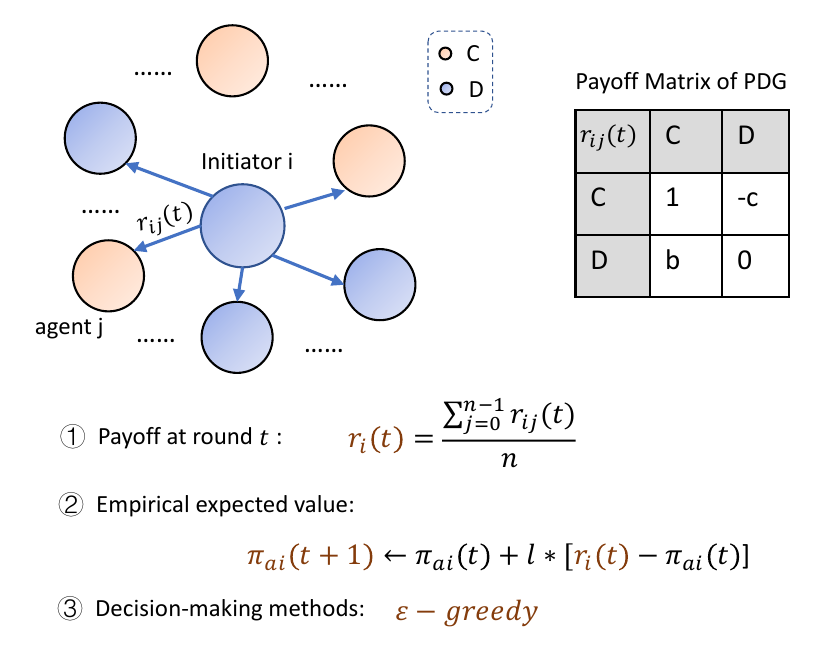}\label{method}} 
    \caption{The schematic diagram illustrates the updating process in the MAS. In round $t$, each individual sequentially assumes the role of initiator $i$ and engages in interactions with $n$ randomly selected individuals, denoted as $j$. The payoff for the current round $r_i(t)$ is calculated using Eq.\ref{eq1}, serving as the basis for updating the experiential expected value for the current strategy. The selection of the next round's choice, $a_i(t+1)$, is determined through the $\epsilon$-greedy algorithm. This algorithm dictates that an individual is more likely to choose a strategy with a higher experiential expected value with a probability of $1-\epsilon$. Simultaneously, there exists a small probability of opting for a strategy with a lower one, thereby embodying a balance between exploration and exploitation in behavior. }
    \label{fig:model}
 \end{figure*}

\begin{enumerate}
    \item In each round $t$, the agent $i$ employing strategy $a_i(t)$ acts as the initiator, where $a_i(t) \in \{0,1\}$, with $0$ representing strategy $C$ and $1$ representing strategy $D$. Subsequently, she engages in games with $n$ randomly selected opponents from the group (denoted as $j$). Based on the payoff matrix of the PDG, the payoff $r_{i,j}(t)$ for each pairwise game is determined. Agent $i$'s average payoff for this round $t$ is then calculated as follows:
    \begin{equation}\label{eq1}
        r_i(t)=\frac{ \sum_{j=1}^{n}r_{i,j}(t)}{n},
    \end{equation}
    which will be used to update her experiential expected values for current strategy $a_i(t)$. While in game theory, ``strategies" typically refer to the set of actions available to an individual, and ``actions" refers to the actual choices made by the individual, in this paper, we do not explicitly differentiate between them, as it does not affect the reader's understanding of the literature. In relation to the experiential expected payoffs of the strategies mentioned below, $r_i(t)$ is also referred to as the immediate payoff for agent $i$ in round $t$. 
    
    \item Secondly, the agent $i$'s experiential expected payoff value $\pi_{a_i}$ will be updated as
    \begin{equation}\label{eq2}
    \begin{aligned}
    \pi_{ai}(t+1) &\longleftarrow \pi_{ai}(t) + l \cdot \left[r_{i}(t) - \pi_{ai}(t)\right]\\
     \pi_{1-ai}(t+1) &\longleftarrow \pi_{1-ai}(t)
    \end{aligned}
    \end{equation}
    When $a_i$ is equal to 0, $\pi_{ai}(t+1)$ actually signifies $\pi_{C,i}(t+1)$, and correspondingly, $1-\pi_{ai}(t+1)$ represents $\pi_{D,i}(t+1)$. Hereafter, we omit the timestamp $t$ of $a_i(t)$ when it matches that of the main parameter. 
    
    The item $r_i(t) - \pi_{a_i}(t)$ is referred to as the prediction error between immediate payoff (actual payoff) and experiential expected value (expected payoff)\cite{feldmanhall2019viewing}. When this error is greater than $0$, the expected value is elevated by the actual payoff, and vice versa. In other words, the experiential value lags behind the actual payoff and gradually converges towards it. 
    
    $l$ is the learning step size \cite{tuyls2003selection}, which allows to adjust the proportion of importance between current and historical rewards in the learning process. In that case, it is in fact an experiential learning method, where
    \begin{equation}\label{eq3}
         \pi_{ai}(t+1) =l*r_i(t) +l(1-l)*r_i(k_1)+...+l(1-l)^{z}r_i(k_z)] ,
    \end{equation}
    assuming an initial value $\pi_{ai}(0)$ of $0$. $k_1$, $k_2$,...$k_n$ are the moments when agent $i$ adopts the same action as $a_i$. When $l=1$, it represents a myopic individual, where agents' next action heavily relies on their recent round's payoff. Despite the formula allowing for an infinite horizon in individual foresight, the discount factor $l \cdot (1-l)^t$ before historical payoffs imparts a forgetting attribute to agents.
    
    \item Eq.\ref{eq2} presents agents' learning method but does not specify which strategy she adopts in the next round. Here, we employ the commonly used $\epsilon$-greedy method from reinforcement learning as the individual decision-making approach, which is a probabilistic method that balances exploitation and exploration. That is,
    \begin{equation}\label{eq4}
        \begin{aligned}
             a_i(t+1)= \begin{cases}\underset{a}{\operatorname{argmax}}\left(\pi_C, \pi_D\right), \text { with } & p=1-\epsilon, \\ \,0 \text { or } \, 1 \quad \text { randomly, with } & p=\epsilon,\end{cases}
        \end{aligned}
    \end{equation}
where $\underset{a}{\operatorname{argmax}}\left(\pi_C, \pi_D\right)$ represents selecting the strategy with the maximum experiential expected value as his action in next round. $\epsilon$ is the exploration parameter, typically ranging from $1\%$ to $5\%$. Under a high probability of $1-\epsilon$, each agent chooses the action with the maximum expected payoff, demonstrating a pursuit of reward maximization. However, there is still a small probability $\epsilon$ that an agent will choose a random action to explore the possibility of higher returns.
\end{enumerate}

The primary modification introduced by this method compared to traditional Q-learning is the shift from learning the relationship between states and rewards to learning the relationship between decisions and payoffs. Consequently, $\pi_{ai}$ becomes agent $i$'s experiential expected payoff for the strategy $a_i$, learned from historical information and serving as a decision criterion with delayed effects. This transformation allows memory to be seamlessly integrated into individual decisions in a concise manner. The simplicity of this approach is evident in two aspects: (1) There is no need to memorize one's own historical payoffs and strategies. The focus is solely on learning their associative relationships. (2) There is no requirement to consider others' decisions and payoffs. It represents a self-learning process under limited information, significantly reducing the algorithm's time and space complexity.

\section{Results}\label{sec3}

\begin{figure*}[!th]
    \centering
   \includegraphics[width=\textwidth]{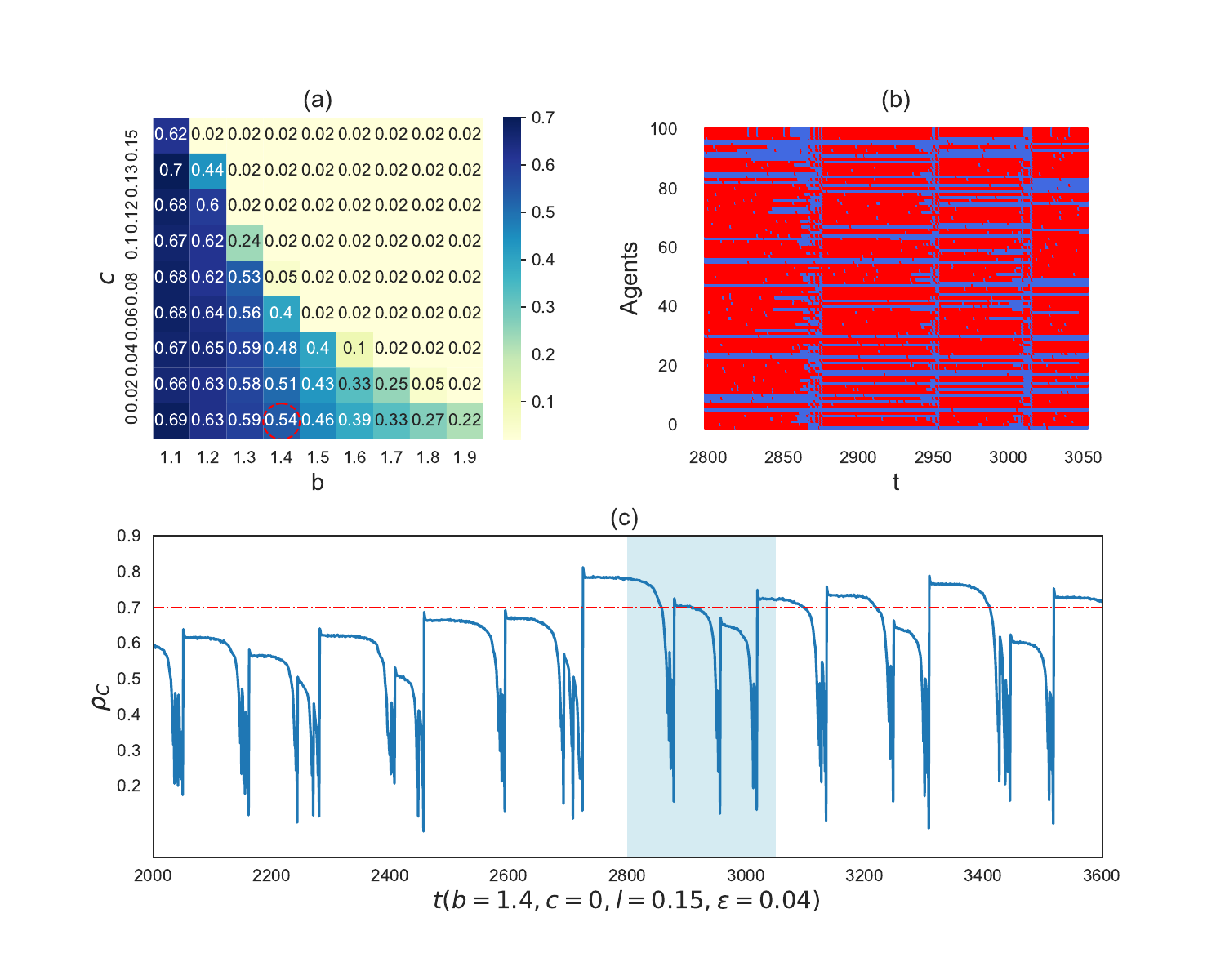}\label{b_c_pc} 
    \caption{Simulation results from a macro perspective are presented as follows. (a) A heatmap illustrates the interrelationships among parameters $b$, $c$, and $\rho_C$, with values highlighted in red circles providing a parameter context consistent with (c). (b) The cell diagram depicts the dynamic changes of the first $100$ agents in the population over the time interval $t=2800\sim 3050$, offering a collective snapshot that corresponds to the blue rectangular depiction in (c). (c) The cooperator frequency $\rho_C$ exhibits periodic oscillations in the well-mixed evolutionary PDG. }
    \label{fig:macro_results}
 \end{figure*}

Our study is based on systematic Monte Carlo (MC) simulations on a well-mixed population of $10,000$. In each MC step, individuals interact with a randomly chosen set of $n$ individuals, obtaining benefits according to the payoff matrix of the PDG. When $n=N-1$, this represents a fully connected structure. After a transitional period, the system tends towards regular changes. The cooperation preference is adopted as a macroscopic evaluation metric, that is,

\begin{equation}
    \rho_C(t)=\sum_{i=1}^N \delta\left(a_i\left(t\right)-0\right) /N
\end{equation}
 When $\delta = 1$ if the initiator is a cooperator at the $t$th step, and $\delta= 0$ otherwise.  
 
Starting from an initial state with an equal fraction of defectors and cooperators, and their $\pi_{ai}(0)$ all equal to 0, we iterate the model with a synchronized update. Simulations are carried out first by varying $b$ and $c$ with $n=N-1$. The results are the average of $20$ trials with various random seeds, aiming to minimize random errors as much as possible. As shown in Fig.\ref{fig:macro_results}(a), the results indicate that even in our model where strategy $D$ is the strict Nash equilibrium point ($c>0, b>1$), the well-mixed population in the PDG exhibits a significantly high cooperative level after evolutionary stability, where $\rho_C\gg 0$. Each data point in the figure represents not the convergence point of $\rho_C$ after the population reaches a stable state but the average value of $\rho_C$ over $1000$ steps after the simulation results stabilize. The actual dynamic evolution of the population cooperation rate over time is illustrated in Fig.\ref{fig:macro_results}(c), displaying substantial periodic oscillations. Combining with the cellular automaton diagram in Fig.\ref{fig:macro_results}(b), we observe that this oscillation of the population cooperation rate corresponds to the alternating actions between order and disorder.

In our MAS, where populations are homogeneous, defective strategies have a clear advantage in the game, and there is no spatial structure, how does such a high proportion of cooperative behavior emerge from disorder?

\section{Analysis}\label{sec11}

\subsection{Group strategy is implicitly included in individual payoff}
\begin{figure*}[!tp]
    \centering
    \includegraphics[width=\textwidth]{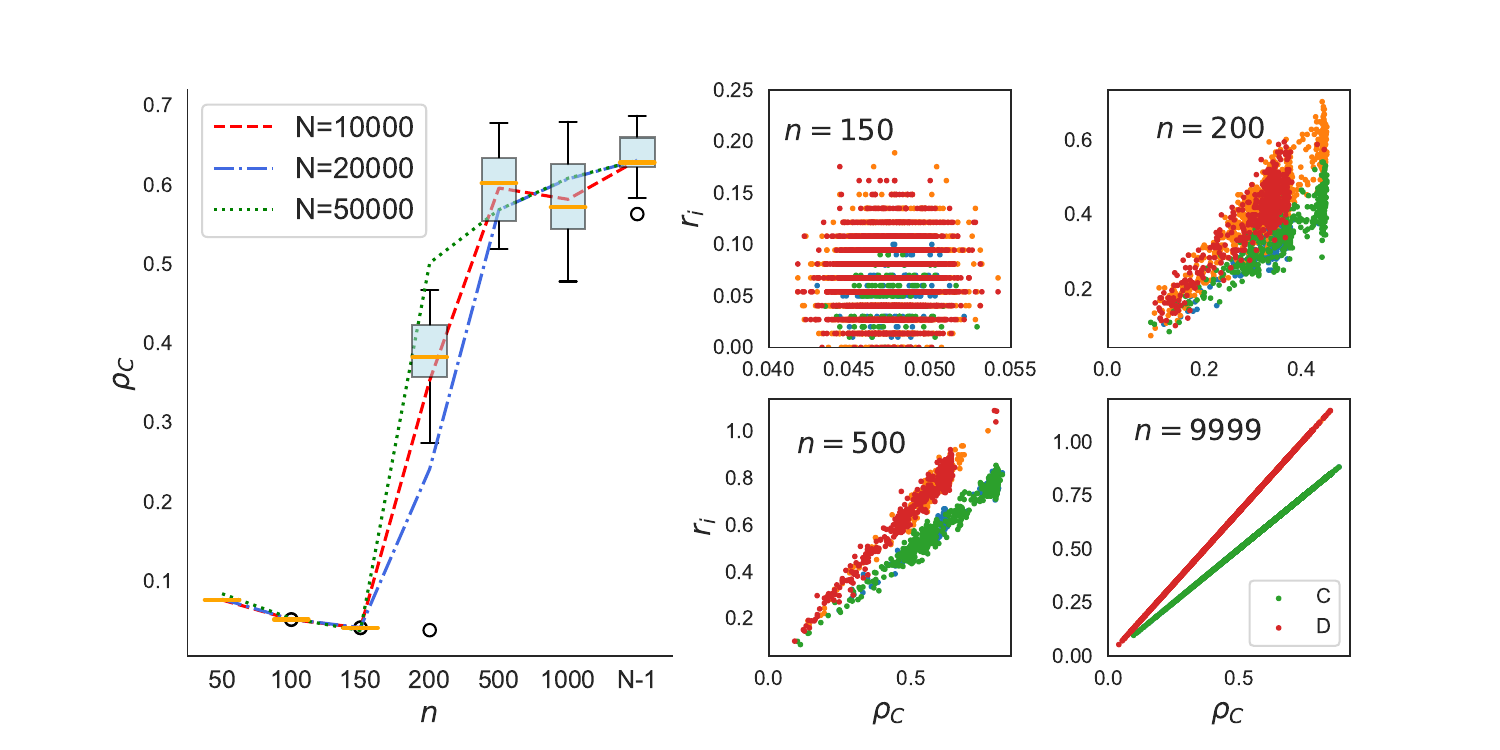}\label{phase_change}
    \caption{The correlation between the number of opponents($n$) that the initiator interacts with and the frequency of cooperators is explored.  (Left) The presence of $\tilde{n}$ is observed; when $n>\tilde{n}$, the frequency of cooperators exhibits periodic oscillations over time. (Right) Scatter plots depicting the rewards of randomly selected agents and the frequency of group cooperation for different values of $n$. It can be observed that when $n<\tilde{n}$, $r_i(t)$ and $\rho_C$ are completely uncorrelated; when $n>\tilde{n}$, this correlation increases monotonically with the increase in $n$, becoming linearly correlated when $n=N-1$.}
    \label{fig:phase_chage}
    
 \end{figure*}

 \begin{figure*}[!tp]
    \centering{\includegraphics[width=0.8\textwidth]{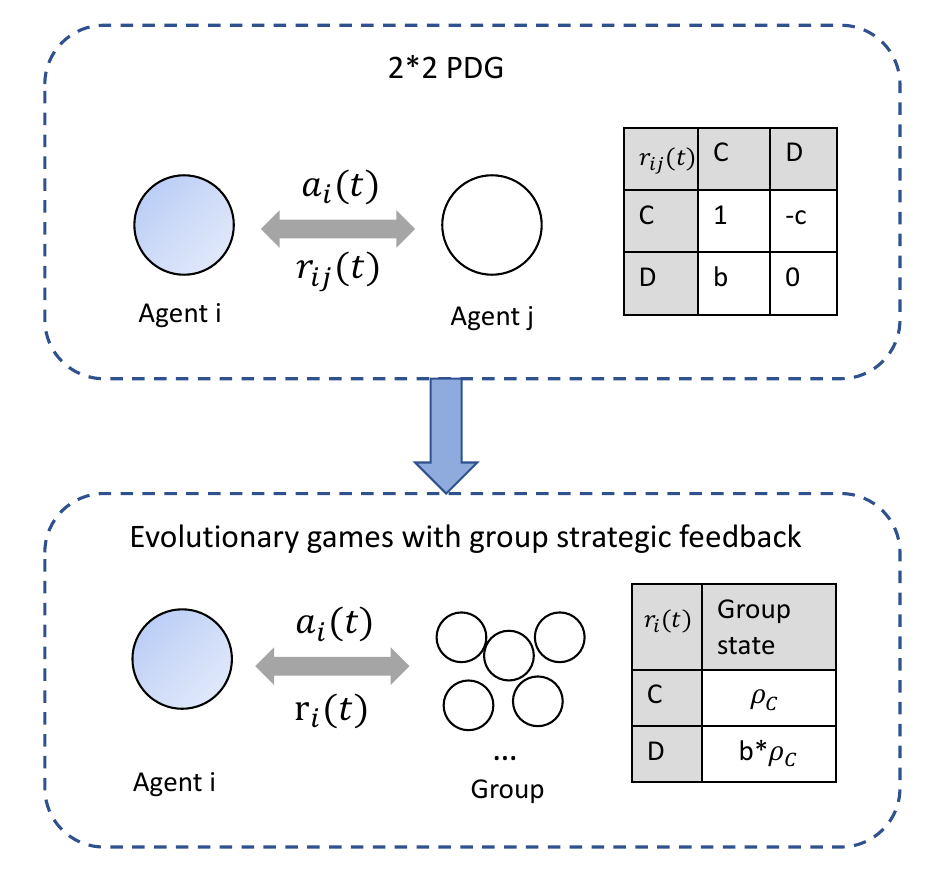}} 
    \caption{When individual rewards exhibit correlation with group strategies, the 2*2 game transforms into a dynamic interplay between individual and group strategies.}
    \label{fig:fig3} 
 \end{figure*}

In our model, agents engage with a randomly selected set of $n$ opponents in the population during each round $t$ and accrue payoffs based on the payoff matrix of PDG. Then, the average payoff is computed for learning. Fig.\ref{fig:phase_chage} elucidates the relationship between the number of opponents $n$ in each round of the game and the average $\rho_C$ after achieving evolutionary stability. As depicted in that figure, a conspicuous threshold $\tilde{n}$ emerges: when $n<\tilde{n}$, the evolutionary outcome favors defectors, and the proportion of cooperators approaches zero. However, when $n\geq \tilde{n}$, the cooperation rate of the population exhibits substantial periodic oscillations.

To explore the reasons behind the emergence of $\tilde{n}$, some agents are randomly selected to examine what information is encompassed in their payoffs $r_i(t)$ under different values of $n$. As shown in Fig.\ref{fig:phase_chage}, when $n<\tilde{n}$, the average payoff per round for individuals is entirely uncorrelated with the proportion of cooperators in the population. However, when $n>\tilde{n}$, a significant correlation emerges, and this correlation monotonically increases with the increase of $n$. When $n=N-1$, $r_i(t)$ is linearly correlated with the population strategic distribution, i.e., the reward of a cooperator is $pc$ and that of a defector is $b*pc$. That is, agent $i$ can derive global strategic information $\rho_C$ from their own round-wise game payoff $r_i(t)$. Under these conditions, the immediate payoff an individual obtains at round $t$ no longer depends on the PDG payoff matrix but rather on the proportion of cooperators in the population, as illustrated in Fig.\ref{fig:fig3}.

The emergence of $\tilde{n}$ can be explained using the Law of Large Numbers. Here, $n$ represents the size of agents' sampling for the group strategy. Let $n_C$ be the cooperation rate in the sampled agents, and $\rho_C$ be the actual one in the entire population. The difference between them is denoted as $E_n= \left|n_C-\rho_C \right|$. According to the Law of Large Numbers, $\lim_{n \to \infty} P(E_n > \varepsilon) = 0$, indicating that when $n$ is sufficiently large, the probability of $E_n$ approaching $0$ increases. In other words, there exists a threshold $\tilde{n}$ such that when $n>\tilde{n}$, the average value $n_C$ of the samples can be employed to estimate the overall average $\rho_C$. Moreover, the average value $n_C$ of the sample will gradually converge to the overall average with an increase in the sample size $n$. In this scenario, the immediate payoff $r_i(t)$ per round for each individual implicitly contains the information of $\rho_C$ . The evolutionary outcome of the well-mixed PDG is no longer the disappearance of cooperators, but rather the presentation of cyclic oscillations between cooperation and defection, as depicted in Fig.\ref{fig:macro_results}. Referring to the replicator equation, the analytical formula for the model can be expressed as following when $n\geq \tilde{n}$, :

  \begin{equation}\label{eq6}
        \begin{aligned}
             \dot{x_C}(t)&=x_C(t)*[1-x_C(t)]*[r_C(t)-r_D(t)]\\
            r_C(t)&=x_C(t)\\
            r_D(t)&=b*x_C(t)
        \end{aligned}
    \end{equation}
  where $x_C(t)=\rho_C(t)$ represents the frequency of cooperators in the population, while $r_C(t)$ and $r_D(t)$ denote the average payoffs of cooperators and defectors, respectively.

Based on the above analysis, we have confirmed that the correlation between the payoff for agents and the state of the population plays a pivotal role in triggering oscillatory patterns among the population. However, we also observe that as $r_D(t)\ge r_C(t)$ holds true, Eq.\ref{eq6} suggests that the proportion of cooperators in the population will gradually diminish until the eventual elimination of $C$ players altogether. The conundrum surrounding the emergence of cooperation remains unsolved. Subsequently, we undertake an in-depth micro-level examination to shed further light on this issue.

\subsection{Memory establishes a coupling relationship between individual and group strategy}

\begin{figure*}[!tp]
    \centering
    \includegraphics[width=1.\textwidth]{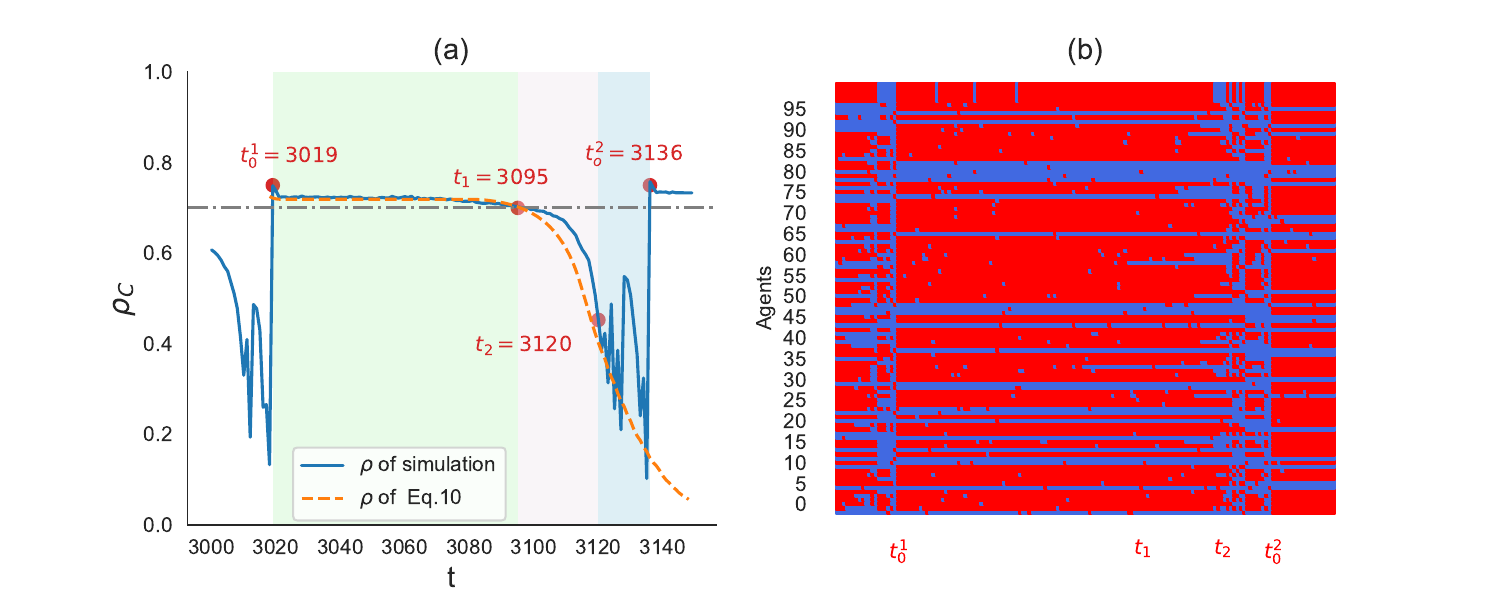}\label{one_cycle}
    \caption{A representative oscillatory cycle ($t=3019 \sim t=3136$) is selected for analysis. (a) The temporal dynamics of the group cooperator frequency exhibit distinct phases: the emergence point $t_0^1$, the stabilization period(highlighted in green, $t_0^1\sim t_1$), the decline period(highlighted in pink, $t_1\sim t_2$), and the disorder period(highlighted in blue, $t_2\sim t_0^2$). (b)The corresponding cellular diagram for this oscillatory period.}
    \label{fig:one_analysis}
 \end{figure*}

As illustrated in Fig. \ref{fig:macro_results}, collective behavior undergoes oscillations between order and disorder, corresponding to the establishment and breakdown of group cooperation. By examining the cellular diagram in Fig. \ref{fig:macro_results}(b) and the population state diagram in Fig. \ref{fig:macro_results}(c), each oscillation cycle can be categorized into distinct phases: the emergence point (where cooperation abruptly forms at a specific time), the stabilization period, the decline period, and the disorder period. To facilitate analysis, we consider a complete cycle from $T=3019$ to $3120$, as depicted in Fig. \ref{fig:one_analysis}, exploring both macro and micro perspectives. Given the instantaneous formation of group cooperation at the emergence point, characterized by clear nonlinear features, our analysis commences with the examination of ordered behavior in the population after this point $t_0$.

\subsubsection{The Stabilization Period of Group Cooperation}

\begin{figure*}[!tp]
    \centering
    \includegraphics[width=1.\textwidth]{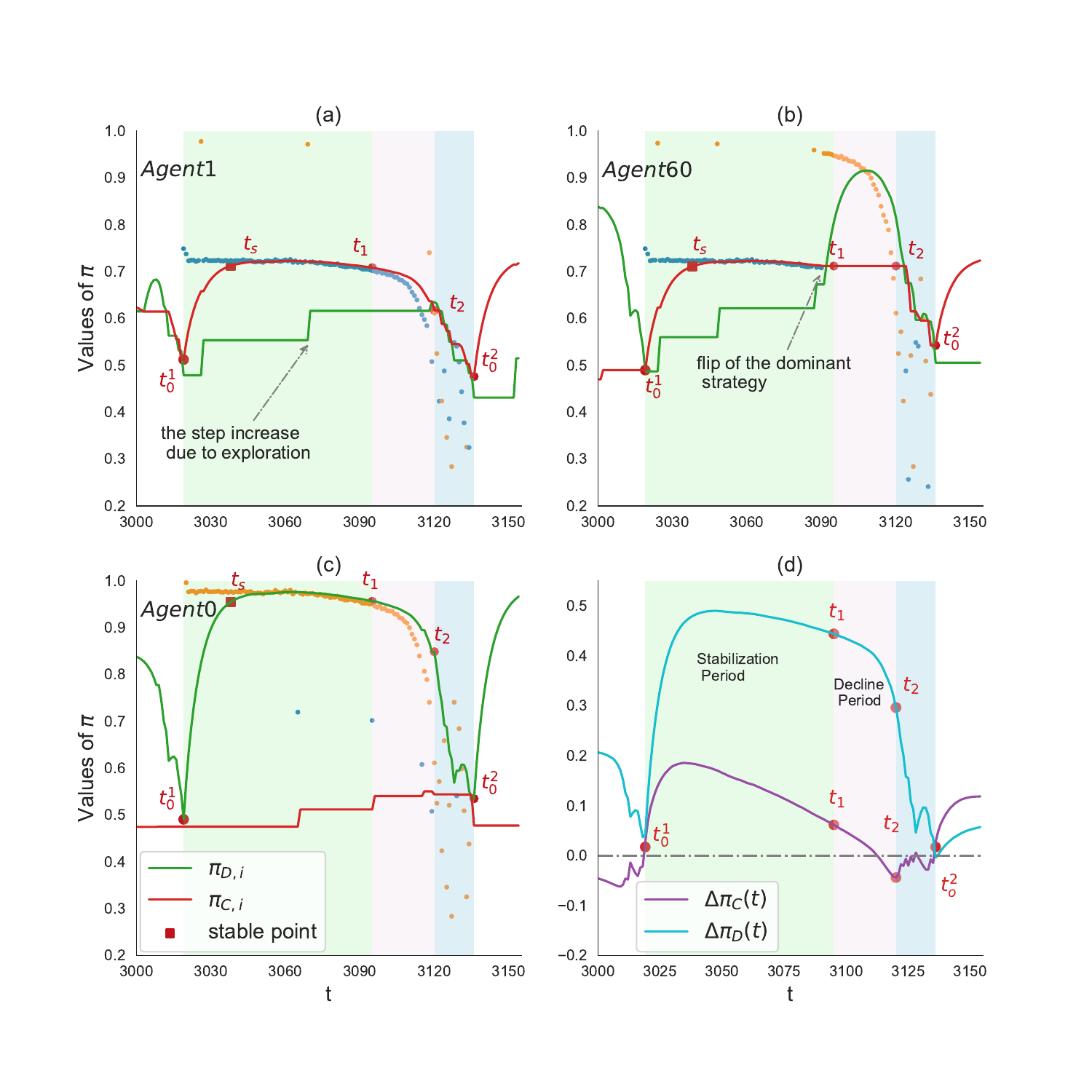}
    \caption{Some representative cooperators and defectors are selected for analyzing the dynamic process of their experiential expected values. (a)(b) Despite agents employing the same learning method, the dynamics of cooperators' experiential expected values vary significantly due to the randomness of the exploration strategy. For instance, $Agent60$ shifts to a defector before the $t_1$ moment, while $Agent1$ remains cooperative until the disorder period. (c) Defectors' experiential expected values remain relatively stable over time, akin to $Agent0$. (d) The disparity between the expected values of different strategies serves as an indicator for the stability of cooperators' behavior and is also used as a criterion to delineate macro-cycles into distinct intervals. In the stabilization period $t_0^1\sim t_1$, $\Delta \pi_C(t)>0$, indicating that the dominant strategy of cooperators remains stable and unchanged (with only a small number shifting to defectors). In the decline period $t_1\sim t_2$, the $\Delta \pi_C(t)$ of cooperators decreases rapidly, even becoming less than $0$, signifying that flipping to defectors becomes the predominant behavior, leading to a swift decline in the group cooperation rate. During the disorder period, $\Delta \pi_C(t)$ fluctuates around $0$, indicating that individuals no longer exhibit a dominant strategy. It's important to note that cooperators here refer to individuals whose dominant strategy is $C$ at the $t_0^1$ moment.}
    \label{fig:indi_analysis}
 \end{figure*}

Some typical cooperators and defectors are selected for comparative analysis from different perspectives. As shown in Fig.\ref{fig:indi_analysis}, the blue dots represent the immediate payoffs when agents adopt strategy $C$, while the orange dots represent that of strategy $D$.  At the emergence point of group cooperation in this cycle $t_0^1=3019$, the population forms a cooperative state with $\rho_C= 0.72$. The experiential expected value of the strategies at that time for agents experience a rapid increase due to the reinforcement of the immediate payoffs. Consequently, agents differentiate between dominant and dominated strategies from the previously alternating actions before $t_0^1$. $Agent1$ and $Agent60$ are thus identified as cooperators (with the dominant strategy being $C$), while $Agent0$ becomes a defector (with the dominant strategy being $D$). 

From $t_0^1$, both defectors and cooperators experience a rapid increase in the experiential expectations of dominant strategies, as depicted during the interval $t_0^1\sim t_s$ in Fig.\ref{fig:indi_analysis}. They eventually converge to the results of Eq.\ref{eq6}, which are $x_C$ and $b*x_C$. In the interval $t_s\sim t_1$, the experiential values of dominant strategies for individuals remain stable as the differences between the experiential values and their immediate payoffs approach $0$. However, the experiential values of dominated strategies gradually increase as agents adopt an exploratory strategy with a small probability $\epsilon$ in each round. This is evident in the step rise of the experiential values for dominated strategies in Fig.\ref{fig:indi_analysis}. Based on the above analysis, the amount of changes in the experiential expectations of dominant and dominated strategies for an average cooperator during this period can be expressed as:
 
\begin{equation}\label{eq7}
    \begin{aligned}
        \dot{\pi_C}(t)&=\pi_C(t)-\pi_C(t-1)=l*[r_{C}(t)-\pi_C(t-1)](1-\epsilon)\\
        \dot{\pi_D}(t)&=\pi_D(t)-\pi_D(t-1)=l*[r_{D}(t)-\pi_D(t-1)] \epsilon
    \end{aligned}
\end{equation}

At the initial time point $t_0$ of each cycle, $\pi_C(t_0)$ approximately equals $\pi_D(t_0)$. Despite the fact that within this interval, the payoff for the cooperators is assuredly much less than that for the defectors, we observe that the frequencies at which these two strategies update are asynchronous, as indicated by Eq.\ref{eq7}. That leads to a significantly greater increase in the dominant strategy, cooperation ($C$), compared to the dominated strategy, defection ($D$). Consequently, the disparity between the two progressively widens. However, starting from the time $t_s$, as the experiential expected value of the cooperative strategy converges, its amount of change $\dot{\pi_{C}}(t)$ approaches zero, whereas $\dot{\pi_{D}}(t)$ continues to rise gradually. This results in the difference between the two beginning to diminish once more. In this context, we can designate,

\begin{equation}\label{eq8}
    \begin{aligned}
        \Delta \pi_C(t)&=\pi_{C}(t)- \pi_{D}(t)\\
    \end{aligned}
\end{equation}
as indicators of the behavioral stability of the cooperator, where a larger value suggests better resilience to disturbances and a lower likelihood of switching strategies. Conversely, the cooperator is less stable and is more prone to strategy reversal in the presence of interference. In the interval $t_0^1\sim t_1$, the stability of the cooperator's behavior undergoes a process of initial increase followed by a subsequent decrease, as shown in Fig.\ref{fig:indi_analysis}(d). However, since $\pi_{C}(t)$ is consistently greater than $\pi_{D}(t)$, there is no significant change in the macro-level rate of group cooperation, which is typical of group behavior during this period. Nevertheless, it is crucial to note that the seeds of betrayal are quietly germinating. With sufficient time, cooperators will eventually discover through exploration that betrayal offers a significant gain advantage over cooperation.

\subsubsection{The Decline Period in Group Cooperation}

\begin{figure}[tp]
    \centering
    {\includegraphics[width=\textwidth]{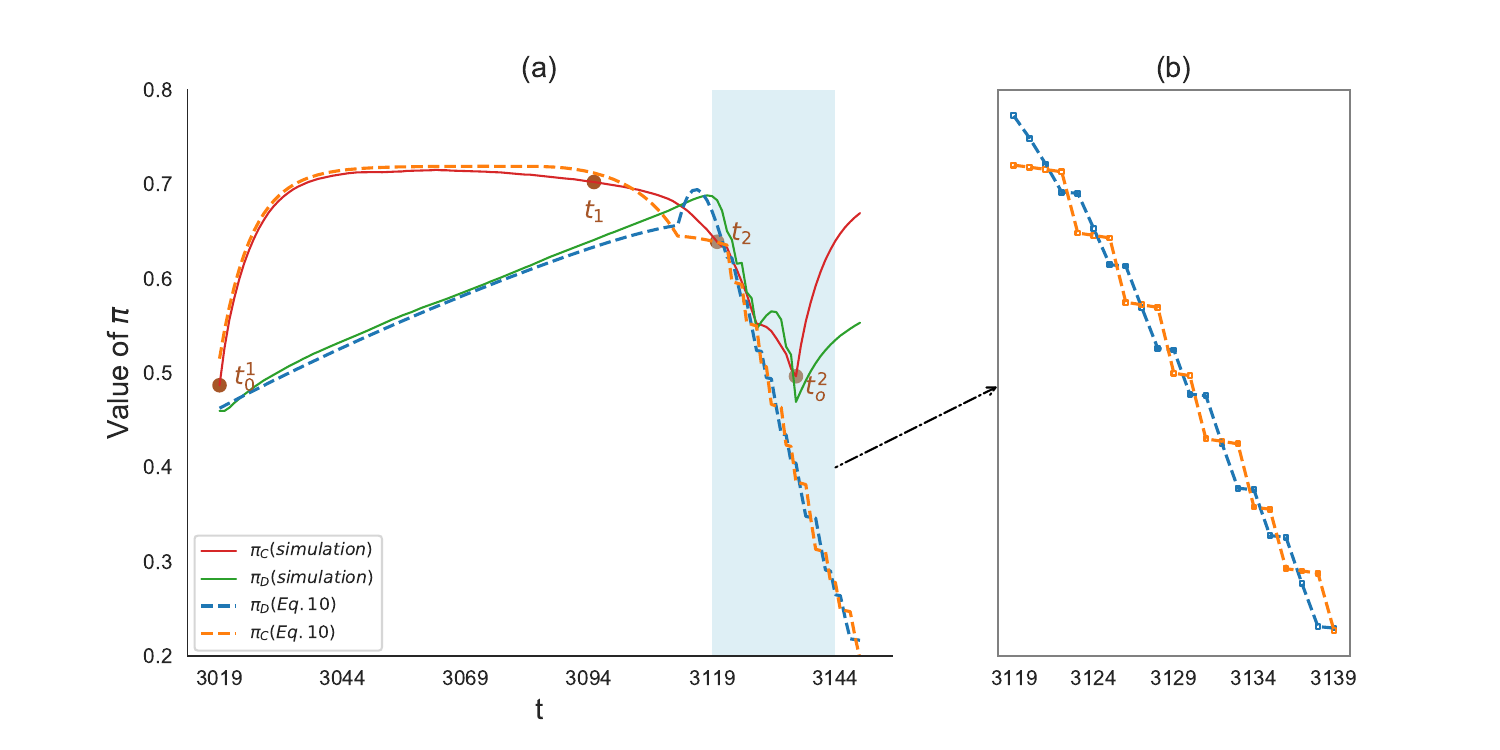}\label{simmulation}} 
    \caption{Comparison of simulation and analytical results for the experiential expectation values of the average cooperator's different strategies during the interval $t=3019 \sim 3136$. (a) The green and red solid lines in the plots depict the simulation results, while the blue and orange dashed lines represent the analytical results derived from Eq.\ref{eq10}. (b) Zoomed-in plots of the analytical results during the disorder period clearly show that different strategies' experiential expectations alternately decrease, indicating that she alternately adopts different strategies }
    \label{fig:simulation}
 \end{figure}

The rise in the experiential expectation of defection during individual exploration leads to a decrease in the $\Delta \pi_C(t)$ of the cooperator. Individuals with smaller $\Delta \pi_{C,i}(t)$ (where $i$ denotes the id number of the cooperator in the group) are the first to shift their strategy, transitioning into defectors, as exemplified by $Agent60$ in Fig.\ref{fig:indi_analysis}(b). The defection of a few cooperators triggers a chain reaction within the group. On one hand, the expectation value $\pi_{C}(t)$ for cooperators decreases. On the other hand, exploratory behavior boosts their $\pi_{D}(t)$, exacerbating the decreasing trend of $\Delta \pi_C(t)$. As illustrated in Fig.\ref{fig:one_analysis}, starting from the moment $t_1$, the rate of group cooperation undergoes a rapid collapse following a gradual decline. More cooperators shift to defectors, and the defective strategy begins to dominate in group. The $\pi_{D,i}(t)$ of these flip-flopping cooperators (e.g., $Agent60$) experiences an evolutionary process of initial ascent followed by decline. This is due to the rapid decline in the rate of group cooperation, causing $\pi_{D,i}(t)$, the dominant strategy, to decrease more rapidly than $\pi_{C,i}(t)$ during this phase. These two values converge approximately to zero at the moment $t_2$ over time, which means those agents enter a random exploration phase. The dynamics of an average cooperator\footnote{All references to cooperators in this section refer to agents whose dominant strategy is cooperation at the beginning moment $t_0^1$ of the cycle.} during this period can be described by the following analytical equation:

    \begin{equation}\label{eq9}
        \begin{aligned}
         \dot{\pi_D}(t)&=l*[b*x_C(t)-\pi_D(t-1)](1-\epsilon)\\
         \dot{\pi_C}(t)&=l*[x_C(t)-\pi_C(t-1)] \epsilon
        \end{aligned}
    \end{equation}

Combining Eq.\ref{eq6}, Eq.\ref{eq7} and Eq.\ref{eq9}, the dynamics of the cooperators' frequency during the ordered period can be expressed as follows:
\begin{equation}\label{eq10}
        \begin{aligned}
        \dot{x_C}(t)&=k*x_C(t)(1-x_C(t))*(\pi_C(t)-\pi_D(t))\\
        \dot{\pi_C}(t)&=l*[r_C(t)-\pi_C(t-1)]*
        [(1-\epsilon)\Theta (\pi_C(t)-\pi_D(t))+
        \epsilon\ \Theta (\pi_D(t)-\pi_C(t))] \\
        \dot{\pi_D}(t)&=l*[r_D(t)-\pi_D(t-1)]*
        [(1-\epsilon)\Theta (\pi_D(t)-\pi_C(t))+
        \epsilon\ \Theta (\pi_C(t)-\pi_D(t))]\\
        r_C(t)&=x_C(t) \\
        r_D(t)&=b*x_C(t)
        \end{aligned}
    \end{equation}

Here, $\Theta(x)$ is the Heaviside function
\begin{equation}\label{eq11}
        \begin{aligned}
            \Theta(x)=\left\{\begin{array}{l}1, \quad x>0 \\
                        1 / 2, \quad x=0 \\
                            0, \quad x<0
                        \end{array}\right.
        \end{aligned}
    \end{equation}

Fig.\ref{fig:simulation} depicts the analytical outcomes of Eq.\ref{eq10}, offering insights into the dynamic evolution of an average cooperator. That closely aligns with the simulation results in the cycle $t_0^1\sim t_0^2$. Through a comprehensive comparison of macro and micro analyses, it becomes evident that when individuals possess memory and continuously learn the correlation between historical payoffs and strategies, the experiential expectations of different strategies are no longer static. Instead, they dynamically change with group cooperation.

During their explorations, initial cooperators discern that defecting offers greater rewards, thereby prompting a shift towards defectors in pursuit of increased benefits. However, as more cooperators transition to being defectors, the overall cooperation rate in the group starts to plummet significantly, reducing the experiential expected value of the defection strategy. This reversal in the experiential expectations between cooperation and defection occurs due to memory introducing a coupled relationship between individual and group strategy. This coupling can erode the payoff advantage of the defection strategy over the cooperative one. Consequently, in the quest for greater benefits, individuals enter a phase of random exploration, propelling the group into a disordered state.

However, at the point $t_0^2$ in Fig.\ref{fig:simulation}, corresponding to the emergence of cooperation in the subsequent cycle, there is a striking divergence between the analytical results and the simulation findings. In the simulations, we observe the onset of another oscillatory period, whereas the analytical predictions suggest an almost complete eradication of cooperators from the system's evolution. This prompts the question: What underlies this discrepancy? And how does collective cooperation emerge?

\subsection{Macro Selection: Unveiling the Emergence of Cooperation}

As shown in Fig.\ref{fig:indi_analysis}(a)(b), starting from $t_2$, as the experiential values of different strategies for cooperators gradually converge, individuals eventually lose their dominant strategies. Simultaneously, with the decline in group cooperation rate, individuals' immediate rewards also significantly decrease, further lowering the experiential values of the corresponding strategies. In other words, whenever a particular strategy is adopted, the experiential value of that strategy tends to be diminished. Therefore, individuals seeking to maximize their benefits will have to alternate between adopting $C$ and $D$ strategies. Whether in the individual graph (Fig.\ref{fig:indi_analysis}(a)(b)) or the analytical graph (Fig.\ref{fig:simulation}), we can observe the serrated decline process of cooperators in the interval $t_2\sim  t_0^2$.

However, a significant disparity between the analytical and simulation results becomes apparent near $t_0^2$. From the analytical results, the group cooperation rate continues to decline, approaching zero (The detailed analysis is provided in Section \ref{secA1}). Conversely, from the simulation results, following a period of decline, the cooperation rate experiences a sudden surge upward, initiating another ordered cycle. This phenomenon is attributed to a synchronous reversal of actions at a macroscopic level during the disorderly phase, where individuals alternate between $C$ and $D$ strategies. The abrupt emergence of a favorable collective state swiftly elevates the immediate payoff for individuals. Consequently, their current actions, reinforced by the high payoff, stabilize and become the dominant strategy. Group cooperation emerges amid the previously disordered dynamics.
 \begin{figure*}[tp]
    \centering
    {\includegraphics[width=\textwidth]{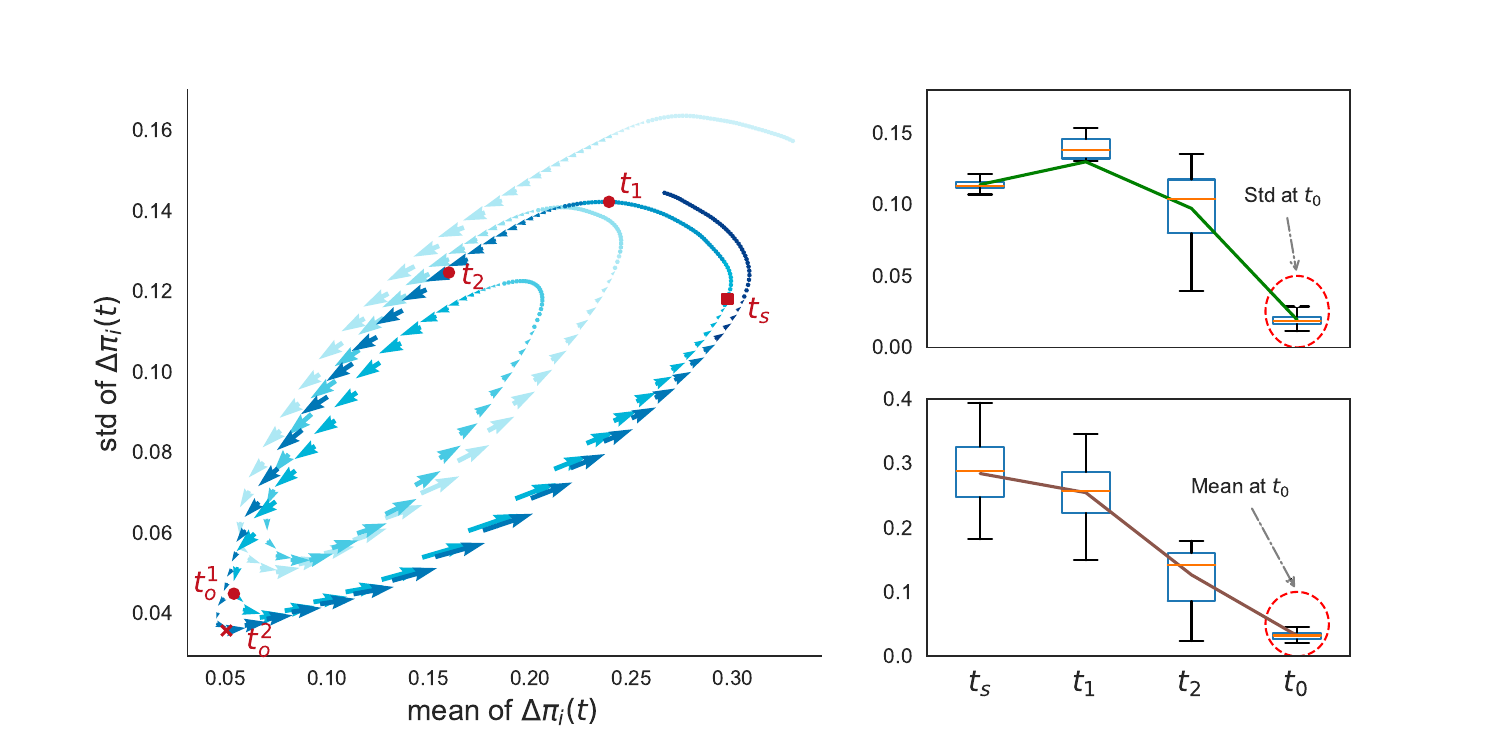}\label{std_mean}}
    \caption{Patterns in the oscillations of collective behavior. (Left) Phase plane dynamics between the mean and standard deviation of the collective distribution of $\Delta \pi_i(t)$ ($i \in [0, N-1]$). Arrows indicate the direction of dynamics over time, and key points align with those in Fig.\ref{fig:one_analysis}. (Right) The distribution of mean and variance of the population's $\Delta \pi_i(t)$ at key points across different cycles. The emergence point $t_0$ is characterized by both the mean and variance being close to $0$. This indicates that each individual lacks a dominant strategy, which leads to maximal disorder within the population.}
    \label{fig:std_mean}
 \end{figure*}

Following Eq.\ref{eq8}, we introduce the indicator of individual stability,
 \begin{equation}\label{individual_s}
    \begin{aligned}
        \Delta \pi_i(t)=|\pi_{C,i}(t)- \pi_{D,i}(t)|
    \end{aligned}
\end{equation}
 
Examining the distribution patterns of $\Delta \pi_i(t)$ within the group, we observe that the emergence of group cooperation exhibits a combination of random and deterministic elements. Fig.\ref{fig:std_mean} illustrates the phase plane, showcasing the variations in the mean and standard deviation of $\Delta \pi_i(t)$ across the population from $t=2500$ to $3500$. The timestamps in this figure precisely correspond to those shown in Fig.\ref{fig:one_analysis}.

Here, we intuitively associate $\Delta \pi_i(t)$ with an individual's confidence in their own strategy. A larger value of $\Delta \pi_i(t)$ indicates stronger confidence in the strategy, leading to more stable behavior and a reduced likelihood of strategy changes. Concurrently, the mean and standard deviation of group $\Delta \pi_i(t)$ at moment $t$ can be interpreted as the group's overall confidence in the strategy and the extent of divergence in this collective confidence, respectively. As illustrated in Fig.\ref{fig:std_mean}, at the moment $t_0^1$, group cooperation emerges, with individuals differentiating into cooperators and defectors. Stimulated by high payoffs, in the interval $t_0^1\sim t_s$, each individual's confidence in their own strategy increases, and the mean of the group's $\Delta \pi_i(t)$ also rises. Meanwhile, the ascending standard deviation indicates an expanding diversity in this confidence, with some individuals becoming steadfast cooperators and defectors, while others are more susceptible to interference and strategy changes. At the point $t_s$, the mean of the group's $\Delta \pi_i(t_s)$ approaches its maximum, indicating highly stable collective behavior. In the interval $t_s\sim t_1$, accompanying individual exploratory behavior, the mean of the group's $\Delta \pi_i(t)$ significantly decreases, signifying a decline in group stability. At the microscopic level, a few cooperators undergo a reversal, turning into defectors. As we progress into $t_1\sim t_2$, an increasing number of cooperators shift to defectors during the exploration process, manifesting as an accelerated decline in both the standard deviation and mean of the group's $\Delta \pi_i(t)$. This trend persists until the moment $t_2$ marks the onset of disorder within the group.

At the moment $t_2$, the standard deviation of group $\Delta \pi_i(t_2)$ is approximately $0.13$, with a mean value of $0.15$. This indicates that not all individuals are alternating between $C$ and $D$ strategies, as there remains a portion of individuals adhering steadfastly to either cooperative or defecting strategies. However, following the disordered phase, both the standard deviation and mean of the group's $\Delta \pi_i(t)$ converge towards $0$, suggesting that nearly all individuals have entered a random exploration stage where they are unsure which strategy is superior. This denotes that the group has reached its maximum level of disorder, and it is under such conditions that cooperation emerges.

It is important to note that, as depicted in Fig. \ref{fig:one_analysis}, during the disordered period, there are brief episodes of order. These periods tend to rapidly collapse due to the low proportion of cooperators. However, amid this trial-and-error process, synchronization occurs in the group's behavioral alternations, meaning that individuals simultaneously switch from defection to cooperation. This synchronized flipping is an integral part of the process that contributes to the emergence of high levels of collective cooperation within the group.

The right subplot in Figure \ref{fig:std_mean} showcases a boxplot that reveals the distribution of mean and standard deviation values for $\Delta \pi_i(t)$ across the population at critical time points within each oscillation cycle ($t\in \{t_0,t_s,t_1,t_2\}$) over $3,000$ rounds following model stabilization. This graphic effectively demonstrates the persistent validity of these evolutionary patterns throughout all cycles.
  The recurring emergence of cooperation is encapsulated by the consistent nature of these collective dynamics. The emergence of group cooperation occurs instantaneously amidst the utmost disordered conditions within each oscillation cycle. Throughout each cycle, individuals experience a gradual increase and subsequent decrease in confidence about their chosen strategies until they reach a state of complete uncertainty regarding which strategy is superior. In this context, different strategies engage in a competition for survival, with the success of strategy $C$ being the result of macro selection. The synchronized shift towards cooperation leads to an immediate surge in individual payoffs, thereby positioning strategy $C$ as the prevailing one. This emergent order within the group arises from this process; however, it progressively disintegrates over time due to individual greed, perpetuating a repetitive cycle of ascent and descent.

The randomness of the emergence of cooperation is manifested in the specific magnitude of the cooperation rate that emerges, the number of individuals that synchronize their behavior, and which particular individuals will become cooperators versus defectors. This randomness makes cooperation emergence an intriguing blend of both deterministic processes and random events.

In summary, as per the analytical formula, the group cooperation rate consistently follows a declining trend since it is determined by the experiential expected values of cooperators and defectors. However, a closer examination of simulation results during this period unveils a fascinating dynamic: despite the overall decrease in the macroscopic proportion of cooperative individuals, at the microscopic level, cooperators continuously relinquish their dominant strategies and engage in an alternating pattern among different strategies. Within this process, individuals synchronize their behaviors at a specific moment, leading to a significant upsurge in collective cooperation. This emergence of cooperative states cannot be discerned from an individual-level perspective, nor can it be analyzed using mean-field theory. It stands as a prime example of how nonlinear dynamics give rise to intricate emergent phenomena that challenge straightforward predictions based on individual components.

\section{Discussion}\label{sec12}

 The exploration of the emergence of cooperation in PDG poses a challenging and captivating endeavor in the field of the evolution of cooperation. The intricacies stem from that game's payoff matrix, where defection stands as the strict Nash equilibrium. Previous studies have predominantly employed methodologies like the replicator equation and the imitator rule to model strategy updating rules among individuals. However, these approaches presuppose that players are myopic decision-makers, primarily motivated by immediate payoffs, resulting in the eradication of cooperators in well-mixed populations. In contrast, our study integrates memory into the strategic decision-making process of each agent and develop a multi-agent model underpinned by introspective learning mechanisms to delve deeply into the fascinating occurrence of cooperation emergence within the evolutionary PDG. By doing this, our aim is to unravel the intricacies of this complex puzzle and potentially reveal new insights into the conditions fostering the development and sustainability of cooperation amid competitive dynamics.

Simulation results illustrate that the integration of memory establishes a coupling relationship between individual and group strategies, leading to oscillatory cycles of cooperation and defection within a well-mixed population. Specifically,  when individuals pursue higher payoffs by adopting a defection strategy, this coupling relationship gradually increases
the number of defectors in the group, causing a decline in their experiential expected payoffs and a loss of relative advantage over cooperation strategies. This compels individuals to alternate between different strategies in pursuit of greater benefits. During this process, when the majority of individuals in the group synchronously
flip from defectors to cooperators, cooperation strategies stand out due to the reinforcement of immediate payoffs. Group cooperation emerges from the disordered state. However, over time, individual greed initiates another collapse of group cooperation, instigating a repetitive cycle where cooperation and defection generate oscillatory dynamics.

The coupling relationship between individuals and the group can be intuitively grasped as a notion of mutual prosperity or adversity, resonating with the aphorism "you reap what you sow." Aktipis et al. have framed this interconnection as fitness interdependence, associating such adaptive reliance with genetic relatedness\cite{aktipis2018understanding}. In contrast to their viewpoint, our study argues that this coupling is rooted in shared interests rather than genetic factors. Building upon this coupling relationship, the gap between individual and group interests is bridged.

Given the simplicity of the model presented in this paper, devoid of spatial structure, noise, and heterogeneity, and eliminating confounding factors, we venture to posit that all mechanisms observed across various studies that facilitate the evolution of cooperation can be traced back to such a coupling relationship. For example, the establishment of direct reciprocity is an outcome of this coupling relationship forming between interacting parties during repeated games. The emergence of indirect reciprocity is not rooted in subjective attributes like reputation or prosociality but rather in individual learning through experience about the coupling relationship between themselves and the group: ``If I help you now, when I need help later, someone will likely help me." Kin selection finds its explanation in an intrinsic genetic-based coupling relationship among individuals. Group selection can be interpreted as ``groups that establish such coupling relationships, forming interest communities, are favored by natural selection, while those that do not are eliminated." Environmental feedback operates on the premise that environment creates this coupling relationship between the group and its individuals, particularly when they need to share public resources. When viewed through the lens of this coupling relationship as a bridge, we find that the theory of natural selection becomes self-consistent, and the conflict between individual interests and group interests depicted by social dilemmas is comprehensively resolved. 

This paper introduces another innovation, the EDM method, which shifts the focus from learning the relationship between states and rewards to learning the correlation between strategies and rewards. Our approach stands out from established memory-based research \cite{wang2016cooperation,liu2010memory,danku2019knowing} due to its simplicity and effectiveness. On one hand, it eliminates the need to store historical rewards and strategies, opting instead to learn their associative relationships, thereby significantly reducing the time and space complexity of the algorithm. On the other hand, it represents a decision-making process based on limited information and self-experience, considering the varying weights of historical and immediate rewards in decision-making, and striking a balance between exploitation and exploration in individual decision processes.

The discoveries presented in this paper contribute to a renewed understanding of cooperative phenomena in nature, shedding light on potential solutions to complex issues in international relations such as the global nuclear crisis, environmental challenges, and food security. Encouragingly, the study reveals that even without any binding force, spontaneous group cooperation emerges among individuals grappling with conflicts between individual and collective interests. This occurs because, metaphorically speaking, we are all in the same boat, and the optimal path to maximizing individual interests lies in cooperation. However, the sobering reality is that this spontaneous cooperation may coincide with a rapid deterioration of the environment or situation, potentially reaching an irreversible and unacceptable state. Consequently, it becomes imperative to explore effective mechanisms, including incentives and penalties, to regulate self-interest, ensure the stability of cooperation, and establish early warning systems for potential collapse. 

\backmatter

\bmhead{Supplementary information}




\begin{appendices}

\section{Analysis of the Decline in the Rate of Group Cooperation by Eq.\ref{eq10}}\label{secA1}

According to Eq.\ref{eq10}, the variation in group cooperator frequency is determined by $\pi_C(t)-\pi_D(t)$. However, during the disorder period, individuals adopt different strategies alternately, and the experiential expectation values of different strategies are successively reduced. In other words, the change pattern of group cooperator frequency alternates between increase and decrease, but overall, the decrease values outnumber the increase ones. Therefore, we can utilize the average experiential expectation value within a $\delta$ step ($\delta\geq2$, ensuring an equal number of updates for different strategies) to estimate the trend of $x_C(t)$, which, according to Eq. \ref{eq3}, is characterized by:
\begin{equation}\label{A1}
    \begin{aligned}
        \pi_{C}(t_2+\delta) &=l*r_C(t_2+\delta -1)*p(t_2+\delta -1) \\
        &+l*(1-l)*r_C(t_2+\delta-2)*p(t_2+\delta -2)\\
        &+...+l*(1-l)^{\delta -1}r_C(t_2)p(t_2) \\
        &+(1-l)^{\delta}\pi_C(t_2)\\
        \pi_{D}(t_2+\delta) &=l*r_D(t_2+\delta -1)*[1-p(t_2+\delta -1)] \\
        &+l*(1-l)*r_D(t_2+\delta-2)*[1-p(t_2+\delta -2)]\\
        &+...+l*(1-l)^{\delta -1}r_D(t_2)[1-p(t_2)] \\
        &+(1-l)^{\delta}\pi_D(t_2)\\
        p(t)&=\begin{cases}
                1,if \quad a_i(t)=0\\
                0,if \quad a_i(t)=1
            \end{cases}
    \end{aligned}
\end{equation}
where $p(t)$ represents the probability of an individual adopting the strategy $C$. Since, during this interval, individuals alternate between adopting strategies $C$ and $D$, it can be approximated that $p(t)=0.5$. Additionally, considering the onset of the disorder period, $\pi_C(t_2)\approx \pi_D(t_2)$. Therefore, in the interval $t_2\sim t_0^1$, we have:

\begin{equation}\label{A2}
    \begin{aligned}
        \pi_{C}(t_2+\delta)-\pi_{D}(t_2+\delta)  &=0.5*l*[r_C(t_2+\delta -1)-r_D(t_2+\delta -1)] \\
        &+0.5*l*(1-l)*[r_C(t_2+\delta-2)-r_C(t_2+\delta-2)]\\
        &+...+0.5*l*(1-l)^{\delta -1}[r_C(t_2)-r_D(t_2)] 
    \end{aligned}
\end{equation}

Combined with Eq.\ref{eq10}, we get
\begin{equation}\label{eq12}
    \begin{aligned}
        \pi_{C}(t_2+\delta)-\pi_{D}(t_2+\delta)  &=0.5*l*(1-b)*\sum_{t=t_2}^{t_2+\delta} (1-l) ^{t_2+\delta -1-t}*x_C(t)
    \end{aligned}
\end{equation}

Due to $b>1$, it is evident that the average experiential expectation values within a $\delta$ step, $\pi_C(t)<\pi_D(t)$. Therefore, $x_C(t)$ will continue to decrease until it reaches $0$.

\end{appendices}


\bibliography{main}

\end{document}